# SEISMOTECTONICS OF BOSNIA - OVERVIEW

Mensur OMERBASHICH [1]* and Galiba SIJARIĆ [2]

[1] *Berkeley National Laboratory, University of California, 1 Cyclotron Rd., Berkeley 94720, CA, USA*
  *Ph. +1-510-486-6780, Fax: +1-510-486-4102,* momerbashich@lbl.gov
  *Now at: Dept of Physics, Faculty of Science, U of Sarajevo. Zmaja od Bosne 35, Sarajevo Bosnia. Phone +387-33-250-483, Fax +387-33-649-359,* momerbasic@pmf.unsa.ba; cc: omerbashich@yahoo.com.
[2] *Faculty of Science, University of Sarajevo, Zmaja od Bosne 33/35, 71000 Sarajevo, Bosnia*
  *Ph. +387-33-250-485, Fax: +387-33-645-328*
*Corresponding author's e-mail*: momerbasic@pmf.unsa.ba; cc: omerbashich@yahoo.com

**ABSTRACT**
Bosnia's seismotectonics seems to follow the Mediterranean marine regime. Earthquakes occur mostly in the outer Dinaric Alps (southern Bosnia), while the strongest earthquakes occur within the Sarajevo Fault system in southern and northwestern Bosnia. In addition to active tectonics being strong, crustal earthquakes occur often as well. Due to Bosnia's rich hydrogeology, crustal loading such as by snow and rain, or reservoir inundation, represents the most important secondary seismogenic source in the region. Despite its exquisite and active geomorphology no comprehensive and reliable geodynamical studies exist on the region. Seismic sensors coverage is extremely poor also. One centenary analogue, and a few recently installed digital seismometers are insufficient for a region that exhibits mild-to-high seismic activity. Significant investments are needed in order for GPS, seismic and other sensor-instrumented networks to be put in place or enhanced. Technical personnel needs to be educated to enable support provide for studies that are done within broader scientific activities. Such efforts that presently seek to include Bosnia under their scope are ESF-COST Action 625, NATO Stability Pact DPPI program, and EUREF/CERGOP geophysics projects.

## GEOLOGY OF BOSNIA REGION

By *Bosnia region* (or simply: *Bosnia*) we mean that part of the Earth's upper crust that encompasses the territory and aquatory bounded by the Sava River in the north, the Una River and the Krka River in the west, the Drina River and the Boka Kotorska Bay in the east, and the Adriatic Sea to the international waters in the south; see Figure 2. We select the midstream portion of the Vrbas River as our *case study* (term used loosely). All seismic magnitudes used in this work are local seismic magnitudes $M_L$, used in the Sarajevo observatory centennial record.

Few contemporary and comprehensive geophysical studies exist on the region, e.g., by Vidović (1974) and by Papeš (1988). Hence the current knowledge on the regional geology is rather sparse. Additionally, due to geomorphologic abundance as well as widespread erosion and vegetation, as well as wind, snow, and other external influences that destroy the surface evidence, only limited reliable information on the Bosnia's geology are available.

According to what is known on its lithofacial development, the region is comprised of various sedimentary, igneous and metamorphic rocks. According to some rough estimates, about 70% of this geologically rich region belongs to the Mesozoic, about 20% can be dated to the more recent Cainozoic, and about 10% to the earliest Palaeozoic eras (Čičić, 2002). Powerful volcanic activity, which largely reshaped the upper crust and ended sedimentation, peaked during the Palaeozoic-Carboniferous. The oldest sediments in the region are the Palaeozoic-Silurian (*ibid*.).

The Dinaric Alps, as a major geotectonic fraction of the Southern Alps makes the dominant tectonic system in Bosnia. Oriented in the northwest–southeast direction, with Bosnia in its central part, this relatively young system laid beneath the Tethys Sea for the most part throughout geological time, until the beginning of the Alpine orogenic cycle that has been going on ever since upper Permian – some 250 million years ago (*ibid*.).

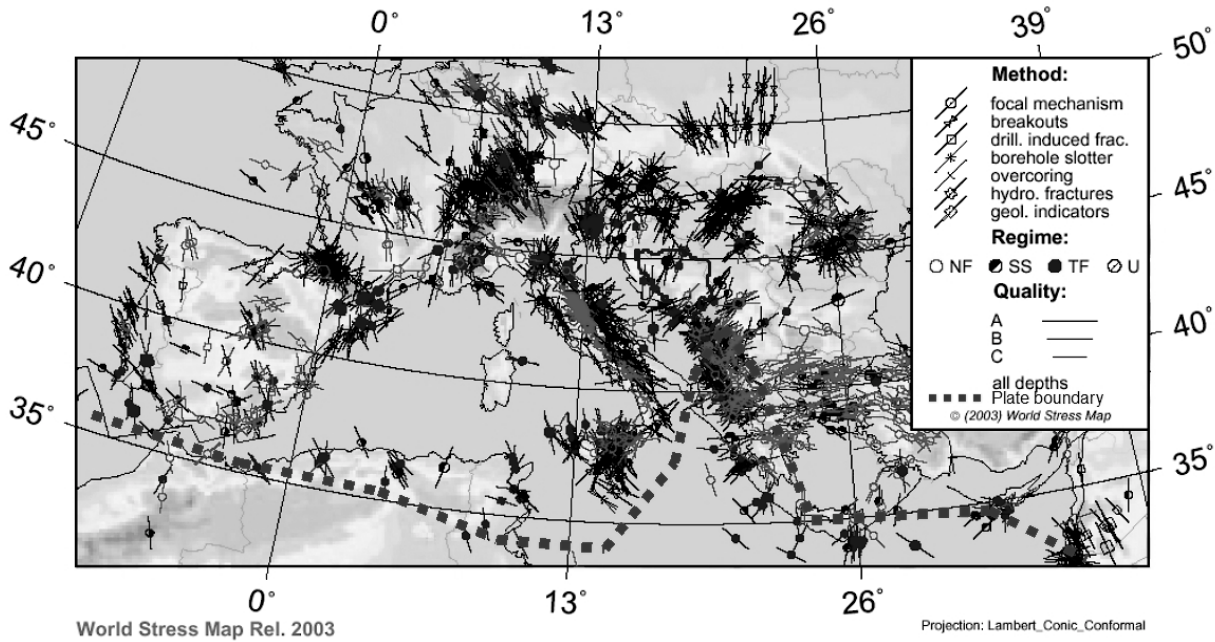

**Fig. 1** Stress Map of the Mediterranean (Reinecker et al., 2003). Shown are horizontal stress directions from different data acquisition methods as part of the World stress map. Dotted line is the trace of African plate boundary. Note lack of data for Bosnia.

Evidence of tectonic activity can be found throughout the region. Thus the karst of the region's south is unique for its large number of closed blind-valley depressions. Blind valleys with flat and alluvial bottoms up to 60 km in diameter can be attributed to active faults as well as intense solution activity in the carbonate rock. The southwest Bosnia is characterized by karst (limestone caves, crevices, and sinkholes), with small depressions containing cultivable soils. Two main passes in the Dinaric Alps lay in the south of the Bosnia region: the Krka River canyon and the Neretva River valley.

Bosnia has substantial supply of a range of minerals, such as iron, bauxite, and coal, as well as rivers of significant hydroelectric potential, that appear however in a geological setup with numerous faulting zones posing a seismic risk due to reservoir inundations, unknown crustal loading effects, and mining. Mining activities in the region had started with formation of first human settlements during the Neolithic Period – between 7000 and 3500 BC. Iron has been mined since $7^{th}$ century BC for export to ancient Greece. Silver, copper, and asphalt were mined throughout the Illyrian era for export to the Roman Empire, mostly for coin minting needs.

**SEISMOTECTONICS OF BOSNIA REGION**

The northward movement of the African plate, and its collision with Eurasia while sliding beneath the European continent, is predominantly influencing the principal horizontal stress directions in the Bosnia region, see Figure 1. Thus the seismotectonics picture of the region is typical of the general marine regime of the Mediterranean belt. This belt is characterized by stages of magmatism and volcanic activity as well as by recycling of sedimentary regions due to collision and subduction. Rheological knowledge derived from terrestrial and space methods points at the same conclusion.

Earthquakes occur mostly in the southern Bosnia, i.e., in the outer Dinaric Alps, see Figure 2. Because of Bosnia's intense hydrogeology, crustal loading due to, e.g., snow and rain masses (Saar and Manga, 2003), and – to a lesser extent – due to large reservoirs (Toppozada and Morrison, 1982), provides the most pronounced secondary seismogenic source in the region. As a result of its tectonics and secondary seismogenic sources, Bosnia sees episodic earthquakes of M5 and above, which are potentially threatening to human life and habitats. Earthquakes that take death toll while causing vast urban destruction have been recorded recently as well, such as the M6.4 Banja Luka event of 27 October 1969, see Figures 3 and 5, preceded by a M4 foreshock of 26 October 1969. The strongest instrumentally recorded earthquake in Bosnia proper over the past one hundred years, as seen in the Sarajevo seismological observatory record, was an M6.5 earthquake of 1923 near a southern town of Tihaljina, Figure 4. Due to the region's general tectonics regime, we focus on deep (25 km or deeper) faults, as they account for the strongest seismicity in the region.

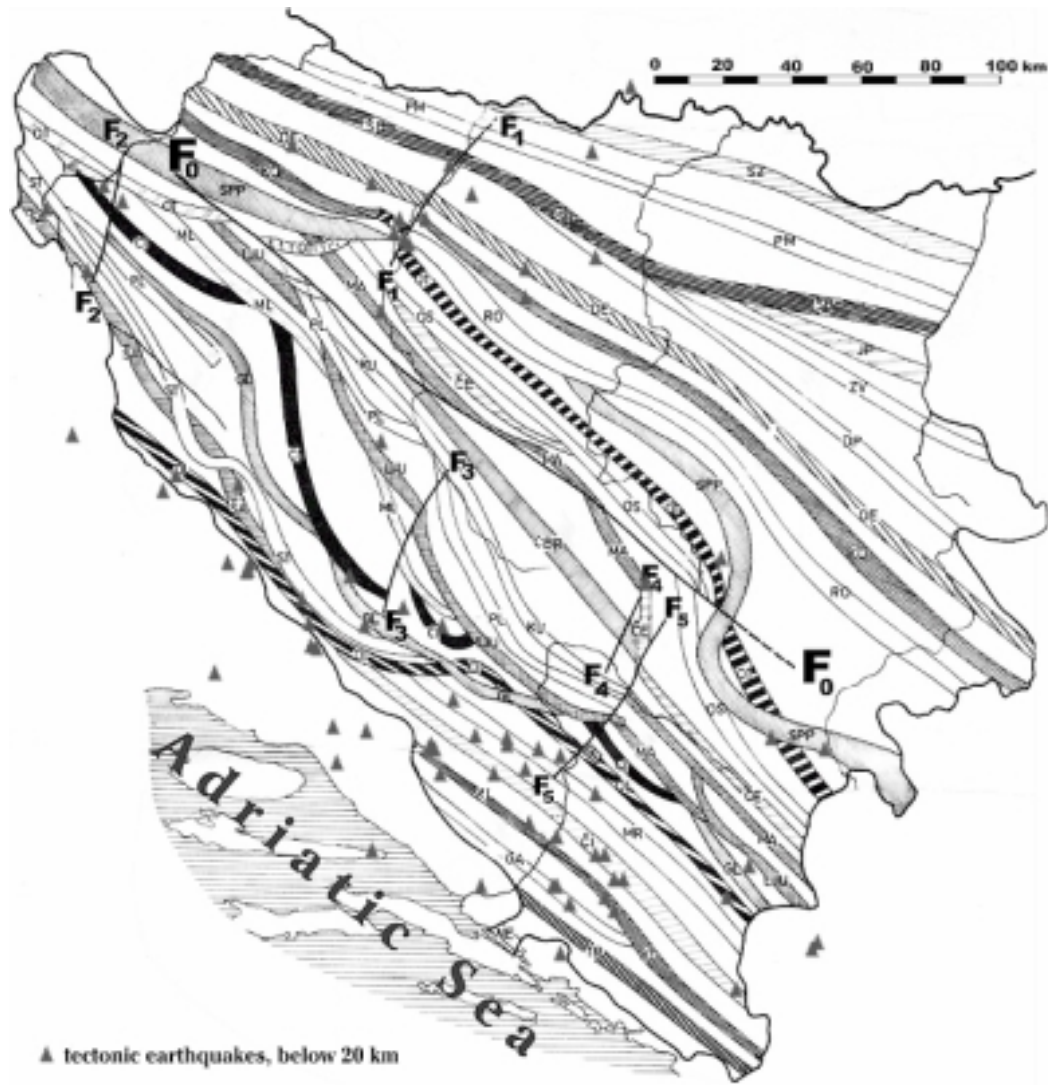

**Fig. 2**  Map of obvious, masked, detected, and presupposed traces of main deep and regional fault system: Sarajevo Fault ($F_0F_0$), Gradiška Fault ($F_1F_1$), Bihać Fault ($F_2F_2$), Livno Fault ($F_3F_3$), Jablanica Fault ($F_4F_4$), and Mostar Fault ($F_5F_5$). Also shown are 30 other tectonic units (distinctive features) represented by different hatches. Map source: Papeš (1988). We plotted on top the epicenter locations (gray triangles) of deep (20 km and below) regional earthquakes that occurred from 1$^{st}$ January 1901 through 31$^{st}$ December 2003, as recorded at Sarajevo seismological observatory. Note the original, Papeš (1988), map showing 30 tectonic units.

Up until the early 1960-ies it was believed that the Bosnian geomorphology is geospatially too complex a system to lend itself for any regularity. Research attempts to date have concentrated mainly on summarizing the surface evidence, and on speculative interpretations of sparse geophysical observations. Thus the first such study – by Vidović in 1974 – had produced a series of feature maps with rather daring ideas. Consequently, and probably in an attempt to give the final word on the issue, some speculative and overly simplistic maps were produced in that study. One such attempt was the *Map of seismogenic zones in Bosnia and Herzegovina* in Vidović (1974, p.118), in which various criteria for establishing geographically the seismogenic potential were lumped together. However, that attempt did not factor in such effects as the nature of crustal loading-related stress redistributions, the rock porosity, etc.

In the most comprehensive study to day of the Bosnia region, Papeš (1988) connected all the pieces of the Western Balkans tectonics puzzle, in a comprehensive and sound manner. As he reported, based on geophysical drilling, electrical measurements, geodetic and seismic surveys, recent deep earthquakes Figure 2, and observational evidence, (*ibid.*), he was able to identify ten complete deep faults and total of thirty tectonic units.

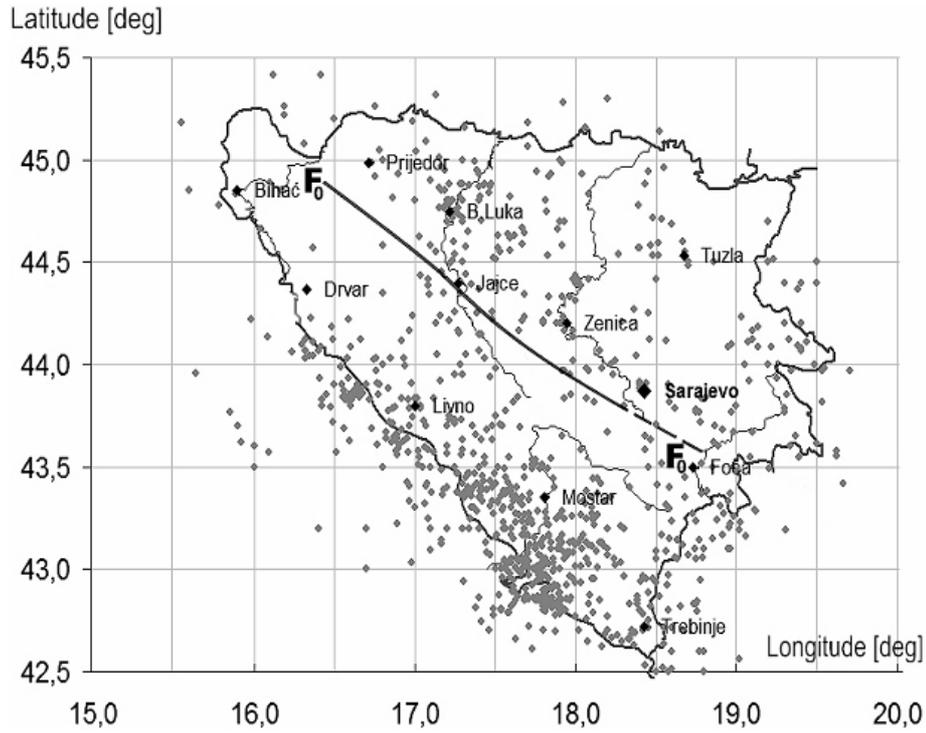

**Fig. 6** Geospatial distribution of region's $M_L$ 2.7+ seismicity (shallow and deep) from historical 1901-2003 record at the Sarajevo Seismological Observatory. Cf. plot of incomplete data provided by Ivan Brlek of Meteorological Service of Federation BiH (Ivan Brlek, personal communication 2003).

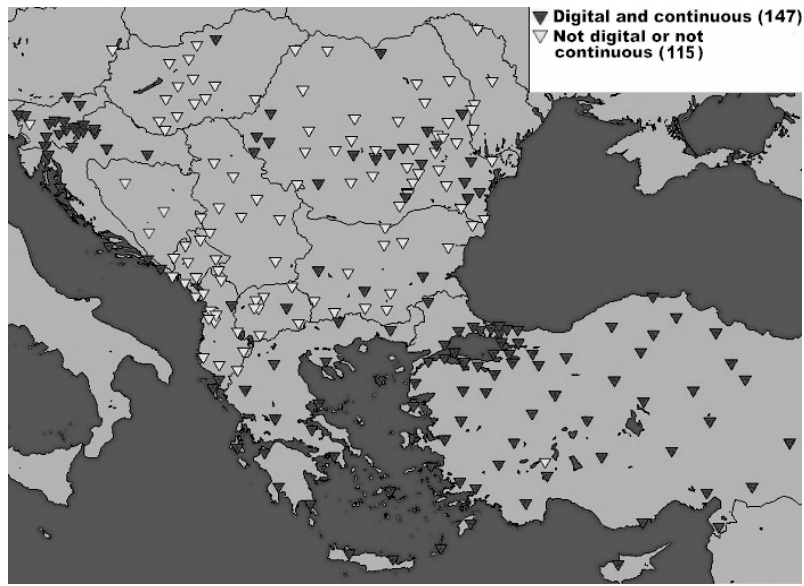

**Fig. 7** Seismological observatories coverage. Updated source: Reinecker *et al.* (2003).

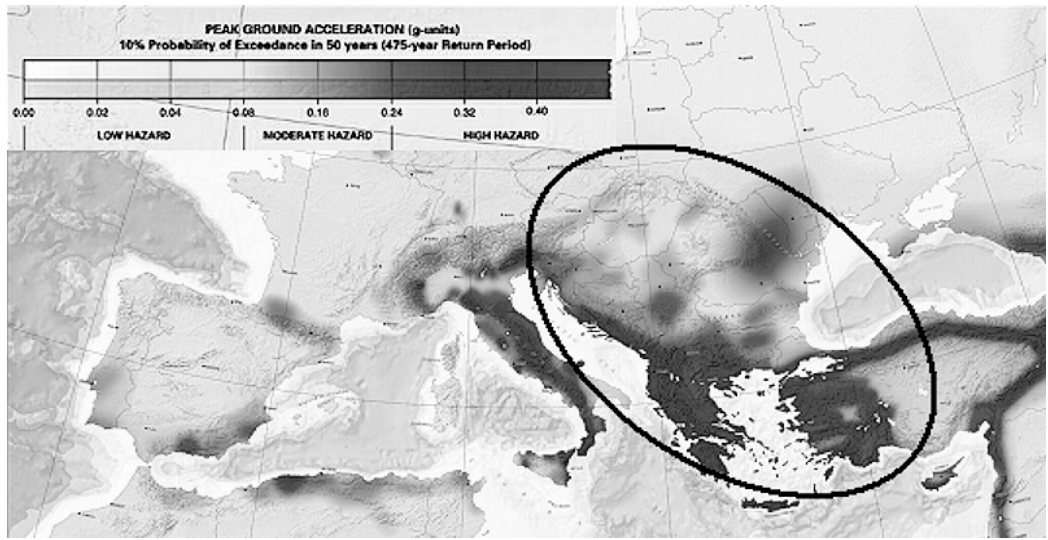

**Fig. 8** Southeastern Europe's Hazard Map, in terms of peak ground acceleration (Reinecker *et al.*, 2003).

Four of ten main (longer than 10 km) deep faults were found as reaching 30-50 km down to the Moho discontinuity. Figure 2 shows the main deep faults in the region. The longest one, the Sarajevo Fault $F_0F_0$, reaches the Moho at 35-40 km, extending over 300 km or the entire length of Bosnia. Other two long deep faults approximately coincide with some of the identified tectonic units, see Figure 2. The second longest deep fault is the Banja Luka Fault down to Moho at 30-35 km, beginning under the (town of) Prijedor Basin west of Banja Luka city, and continuing in the west-east direction while coinciding with the traces of the Kladanj (KD) and Devetak (DE) tectonic units. These units encompass such surface thermal manifestations as the Olovo, Kladanj, and Višegrad spas. The Banja Luka Fault then finally exits underneath the Drina River on to Serbia. The third long deep fault is the Konjic Fault, reaching the Moho at 40-45 km, originating off of the Sarajevo Fault underneath the town of Jajce, and coinciding with the traces of the Kupres (KU), Manjača (MA), Ljubuša (LJU), and Glamoč (GL) tectonic units that contain gaseous water wells, and finally exiting on to Montenegro. Main transverse deep faults are the Gradiška Fault ($F_1F_1$), the Bihać Fault ($F_2F_2$), the Livno Fault ($F_3F_3$), the Jablanica Fault ($F_4F_4$), and the Mostar Fault ($F_5F_5$), Figure 2. All main transverse deep faults submerge under the Sarajevo Fault (Papeš, 1988), while the Livno Fault reaches the Moho; the depth of the M4.7 earthquake of 23 May 1974 beneath the town of Livno was estimated at 38 km. Obviously, depth estimates in Papeš's work were largely severed by sparse seismometer coverage.

There are indications however that other deep fault systems exist in the region besides the ones mentioned herein, and besides the known Sava River Fault in the north. For instance, the M4.3 earthquake of 8 April 1989 near the town of Trebinje, 25 km east of Dubrovnik, was estimated at 39 km depth; as well as other deep earthquakes and deep sequences in the same area, see Figure 2. The Bosnia's deepest recorded event was the M3.7 earthquake of 26 May 1997, underneath the town of Lištica at the southern end of Mostar Fault, estimated at 47 km. The deepest earthquakes that occurred over the past century were mostly located around the towns of Livno, 6 of 73, Bihać 4 of 8, and Bosansko Grahovo, 4 of 22. An interesting sequence occurred near the town of Bosansko Grahovo between 25 November – 24 December 1986, when four earthquakes of M5.9, 3.6, 4.4, and 4.9 had stricken at depths of 21, 13, 15, and 29 km, respectively. Strong sequences can be found throughout the Sarajevo observatory centennial record. Geospatial distribution of region's $M_L$ 2.7+ seismicity from this historical 1901-2003 record is shown on Figure 6.

Centennial deep seismicity as plotted on Figure 2, suggests that faulting systems in the region are comprised mostly of mature faults of substantial strength. High seismic activity along the transverse deep faults, in contrast to low-to-mild seismic activity along the major (longest) deep fault zone, indicate that the Sarajevo Fault $F_0F_0$, as the longest deep fault system in Bosnia, as well as the $F_1F_1$ Gradiška Fault Figures 3 and 5, as a branch of $F_0F_0$, could have a high strength and therefore episodes of M6 earthquakes or stronger could be expected in their locality. (Note however a lack of information on very strong earthquakes in the recent geological past of up to 50,000 years ago.)

The last century's strongest, Tihaljina earthquake of 1923, took place in the locality of the intersection of two deep faults, some 50 km southward along the extension of the Livno Fault $F_1F_1$, and 30 km southward along the extension of the Mostar Fault $F_5F_5$, see Figure 4. Recently, as it is generally expected from long such faults, M6+ earthquakes occurred around the midsection of the Sarajevo Fault at the proximity of the Gradiška Fault – in 1888, 1935, 1969, and 1981, between the cities of Banja Luka and Jajce. Thus it would appear that a strong earthquake of M6 or above is overdue all through the midsection of the Sarajevo Fault joint with the Gradiška Fault.

Bosnia's seismogenic sources may seem akin to those of other regions in the world. Comparing a fault system in one area of the world to a fault system in another part of the world can enhance knowledge on either of the fault systems spatial-temporal evolution; see, e.g., Tondi and Cello (2003). The Sarajevo Fault traverses the entire length of Bosnia. It is generally recognized that long faults suffer most stress around their midsections, while stress drops toward the fault's ends. As mentioned above, this also seems to be the case with the Sarajevo Fault $F_0F_0$, see Figure 2. Furthermore, southern Bosnia experiences high seismic activity albeit of mild-to-high intensity. It is traditionally viewed as helpful that frequent earthquakes seemingly aid the faulting system to release a portion of its stress, nevertheless mostly at fault's ends, or in case of Bosnia around the main fault's branches Figure 2. On the other hand, the central part of the Sarajevo Fault sees strongest earthquakes not as frequently but apparently episodic. (Note that the Sarajevo Fault is not seen as a classical strike slip fault.) Finally, it can be stated that, relative to the general seismicity regime of Bosnia, the Sarajevo Fault is the strongest and in terms of strong earthquake occurrence one of the quietest faults in Bosnia.

As previously reported by many researchers, new reservoirs that are impounded across the midsections of large faults can cause the total fault stress to exceed a normal strength of the fault. Leith *et al.* (1981) give one such account of a connection between local structure and regional tectonics. A newly proposed Vrbas River reservoir near the city of Banja Luka lays central to the Sarajevo Fault. The proposed lake supposedly should encompass the Grebenska Cliff (*Grebenska klisura*) in the length of about 30 km, see Figure 3. It is possible that a nearby, recently built reservoir Bočac (pronounced: botch-utz) has already induced an unspecified increase in the level of low-to-mild seismic activity in the same area all the way to some 50 km upstream the Vrbas River.

The proposed accumulation lake, see Figures 3 and 5, would inundate the section of the Vrbas River in the Grebenska Cliff between the town of Bočac and the city of Banja Luka, of about 30 km in length, and with varying water mass volume, i.e., varying recharge phase cycles of reservoir level. The proposed reservoir's bedrock composition includes mostly karst (limestone and dolomite) conglomerates and marl, with sediments in the midsection, see Figure 5. This case scenario can lead to the reservoir-induced, or porosity-related seismicity (Chen and Talwani, 1998).

Moreover, the unknown stress redistribution effects, combined with landslides and pore pressure-related and snowmelt-related deformities of the upper material, are likely to be present throughout the region. The proposed Vrbas River project could thus result in series of karst or even tectonic earthquakes along the $F_1F_1$, Gradiška Fault alone, which already saw episodes of M6+ earthquakes, see Figure 2. Besides, we saw that the Gradiška Fault already belongs to the earthquake-prone midsection of the calmest (and presumably strongest) Sarajevo Fault $F_0F_0$.

In addition to Bosnia's specific tectonic setup that includes rich hydrogeology, there is a significant level of mineral resources exploitation going on in the presence of karst caves. Such excavation in Bosnia typically goes on for several centuries to several millennia, and occurs often in the proximity of reservoir-inundated areas. The bauxite ore mines around the Grebenska Cliff, see Figure 5, pose one such risk. This adds to the hazard from secondary seismogenic sources in the region.

Of 262 seismic stations in Southeastern Europe, see Figure 7, only 3 operate in Bosnia, of which two digital but not continuous stations are located in the cities of Banja Luka and Mostar since a few years ago, and one analogue station in Sarajevo since 1899. Figure 8 shows a current seismic hazard map of Southeastern Europe in terms of peak ground acceleration at 10% probability of exceedance in 50 years on 475-year return period (Reinecker *et al.*, 2003). It can be seen from Figure 8 that Bosnia – its southern part in particular – is positioned in a high seismic hazard zone.

**GEODYNAMICS STUDIES OF BOSNIA REGION**

Available studies on the region's geodynamics are very sparse. Previous field campaigns seldom sought to determine the dynamics of the region as a whole. One regional study by Jovanović (1967) used centennial precise-leveling data to establish the recent crustal uplift in the region. Data from precise-leveling campaigns of 1931 *vs*. 1873, and 1961 *vs*. 1946, were used in *ibid*. He found the uplift be in the order of some mm/yr; see Figure 9. This is in agreement with the general tectonics tendency in the region, see Figure 1, as well as the subsequent study by Papeš (1988).

Modern geodynamical studies also concur with the above findings on crustal uplift. Mostly northwest-striking faults Figure 1 indicate horizontally that there is a fault trend coincidental with the Dinaric Alps. Unfortunately, modern geodynamical studies that used GPS to detect potential horizontal movements such as the so-called lithospheric compression and expansion, have failed to produce regionally meaningful results, owing to low spatial density of GPS stations used in those studies. The permanent GPS station "SRJV" in Sarajevo is the region's only permanent GPS station. It is a part of the Central European GPS Reference Network (CEGRN). Preliminary results of ~5 mm/yr northern-to-northeasterly tendency, as per unpublished CEGRN vector-solutions using the SRJV half-decade long record, point at the above general conclusion on the northern-to-northeasterly trends of the Dinaric Alps as a whole.

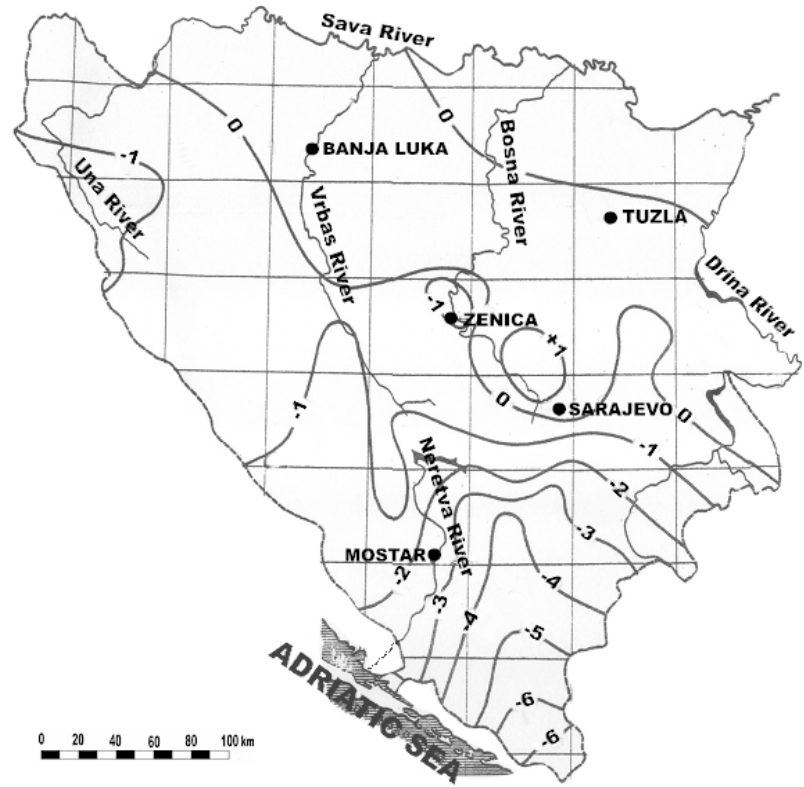

**Fig. 9** Recent crustal uplift velocities in mm/yr, from centennial precise-leveling records. Horizontal grid increment: 30' longitude. Vertical grid increment: 20' latitude. (Jovanović, 1967).

The Bosnian only permanent station "SRJV" is situated at the roof of the Department of Geodesy at the University of Sarajevo. The 5° elevation mask was selected. A PC station on uninterrupted power supply is part of the installation. The instrument is a Trimble SSI 4000 receiver without any auxiliary gauges, which are to be added. The station became operational on 11 June 1999 at 17:14 UT. It operates nearly continuously, with nightly FTP uploads, and certain interruptions. "SRJV" station contributes to the European weekly GPS solution. Geotechnical characteristics of the underlying bedrock primarily include clastic and marl Miocene Tertian bedrock type, with some clay. In addition, the host building is situated on a terrain that is rather sloped in the southward direction, and lies in a close proximity of former extensive depots of brick material. This is far from a perfect choice of location for a tectonic observation station; the economic considerations however were detrimental in making this specific selection. For the above reasons the true velocity could be greater for the "SRJV" station than preliminary suggested. Use of GAMIT/GLOBK software is planned for future works.

The present coverage of Bosnia by GPS campaigns is somewhat satisfactory. The state networks seem dense enough to serve as a basis for fostering geodynamics investigations. This still needs further efforts so as to improve on the lack of spatial-temporal seismotectonics information. Forthcoming GPS densifications are to fill spatially the gaps in the present GPS-stations coverage, at some of the most important rheological/hazardous features such as the Livno Fault, the ever sinking Tuzla salt mine system (one of the Europe's largest), and so on. The upcoming densifications of the permanent GPS networks so as to include at least two such new GPS points should further enhance the resolution of vector-solutions. Moreover, such densifications should enable an overall hazard assessment in view of the improved spectral analyses of the 1901-present centennial historical record from the Sarajevo seismological observatory. This record contains valuable information on local and regional seismicity, and it is one of the Europe's oldest and most complete such instrumented archives.

The State Geodetic Surveys of Bosnia have conducted total of seven GPS campaigns, with 36 GPS-observed stations so far. These 36 stations were observed primarily as an aid in assessing the overall geospatial sparseness in the region. Those campaigns included the "CROREF 1996" and "BIHREF 1998", both under the auspices of the European Reference Frame (EUREF) project, the "Extended SAGET 1998" within the European Satellite Geodynamical Traverses project, as well as the Consortium for Central European GPS Geodynamic Reference Network (CEGRN) "CEGRN 99", "CEGRN 01" and "CEGRN 03", all within the CERGOP-Central Europe Regional Geodynamics Project. There was also one regional densification campaign, "BIHREF 2000".

So far, Bosnian GPS campaigns have each included five static 24 h sessions at 15 s sampling intervals, under a 15° elevation mask, where L1/L2 wGP antennae were used. Data were processed at BKG/IFAG Frankfurt using *Bernese* software v.4.0. The unpublished CERGN coordinates standard accuracy is 2.0 mm NE, 6.5 mm in height.

**DISCUSSION AND CONCLUSIONS**

Since geodynamics of the Bosnia region is not sufficiently understood, it is of key importance to determine the strength of revealed and presupposed faulting systems, namely the Sarajevo Fault, in order to classify creep and slip faulting. Prioritizing and subsequently creating new regional and local seismic hazard maps is called for, also.

For that purpose, geodynamical GPS studies of friction force and normal stresses in fault systems are necessary. This would build on the current knowledge gained from 3D geodynamical GPS studies that are rapidly becoming substantial. Structural analyses and morphology studies ought to be undertaken, also; see, e.g., Borre et al. (2003). The rock pore pressure and cohesion must be determined; see, e.g., Townend and Zoback (2000), as well as the role of bedrock structure on the distribution of elevated pore pressure and reservoir induced seismicity; see, e.g., Talwani (2000). Fault maturity and gauges have to be recognized from such new geological field campaigns. This is particularly important given the fact that most of the Bosnia's main faults do branch into many transform fractures along which most of the region's seismicity occurs, see Figures 2 and 6. New campaigns should therefore aim at identifying the most active main faults and fault systems in the region. Since Bosnia is a marine country, we recommend using the mean sea level for the definition of the horizontal reference plane in determining the faults' characteristics.

Bosnia has numerous deep and active, hence seismogenic faults. New studies as well as study revisions are needed to determine heat flow conditions beneath Bosnia, and to examine earthquake swarm processes and their potential for reactivating large inactive main faults, particularly in the northeast and in the south. Such potential could be also expected for reservoir loading that could change the stress distribution within the region which probably has not experienced catastrophic M7.5+ earthquakes over the last several millennia. To complete such studies, the depth of major faults, i.e., those capable of producing damaging earthquakes, should be re-established from existing records, as well as from new field studies too. The Tihaljina M6.5 earthquake of 1923 should be reassessed; see, e.g., Cello et al. (2003). Spatial clustering tendencies as inferred from Figure 6 for the Banja Luka and the Livno faults, both in the midsection of the Sarajevo Fault system, should be examined further; see, e.g., Bala et al. (2003).

Since only geospatially partial historic records, extending over a century or so, exist in the region, and given the fact that triggering mechanisms for reservoir loading are largely unknown, our remarks on the Vrbas Reservoir ramifications should be taken with caution. However, even if the geological setup were not as grave as it may appear in the case of the Vrbas Reservoir project, a permanent telemetric network of seismometers with at least 5 to 10 sensor stations in the vicinity of the reservoir should be made available for any such project. Some other measures that are a must in that project include the adjusting of inundation phase cycles so as to avoid the summer periods. This should be done because of the possibility of seasonal groundwater/snow recharge-induced seismicity as the area receives huge amounts of precipitation/snow; see, e.g., Saar and Manga (2003). The duration of inundation phase cycles should be kept shorter rather than longer, since longer cycles of around 1 year in duration tend to be correlated with the occurrence of larger and deeper earthquakes, according to, e.g., Talwani (1997).

The current state of affairs in geodynamics studies in the region is promising. Auxiliary GPS studies should aim at establishing denser GPS control networks. Other geophysical fieldwork is well overdue, such as new precise-leveling campaigns and new gravity surveys, followed by seismic refraction surveys, electric resistance profiling, laser distance measurements of geodetic baselines, geomorphology studies, river terrace inclination surveys, gathering of relevant satellite data, and so on.

Initiatives to revive geosciences research in Bosnia include the NATO Stability Pact DPPI program in seismology, and the European

Commission (EC) / European Science Foundation (ESF) COST Action 625 – 3D monitoring of active tectonic structures. Initiatives to revive geosciences teaching in Bosnia include EC TEMPUS projects. Ongoing projects to foster geosciences research in Bosnia include the EUREF (European Reference Frame) project in geodesy, and the CERGOP (Central Europe Regional Geodynamics) project in geodynamics. Outcomes of to-date activities in some geosciences in Bosnia are at somewhat satisfactory level, e.g., with a rising international cooperation in structural geology, GPS geodesy, as well as in protection of Bosnian geoheritage through the European ProGEO initiative. These activities make for a solid foundation to build on.

The next steps should follow modern trends in equipment and methodology. Substantial lack of equipment as well as of qualified personnel had prevented any serious attempts in seismology, too. Dozens of new seismic sensor stations are needed. Due to lack of adequate (in the past decade – any) material support, most geological studies to date, including the Papeš's work shown here, represent a summary of sparse knowledge on the region's geological past. Substantial need exists for speedy recovery of geosciences in Bosnia in general – primarily in data acquisition and analysis, as well as in the equipment and software. International cooperation should help enhance training, and instate *exchange* as the keyword.

**ACKNOWLEDGEMENTS**

Authors thank Dr. Luigi Piccardi of Camerino U., Italy, for his useful suggestions as well as his enthusiasm in including Bosnia into current scientific activities within European integrations. Mr. Ivan Brlek of Meteorological Service of Federation BiH provided Bosnia's historical earthquake records/plots. This work was enabled in part by Ministry of Education and Science of the Sarajevo Canton, COST Action 625, as well as Tempus project CD-JEP-16145-2001.

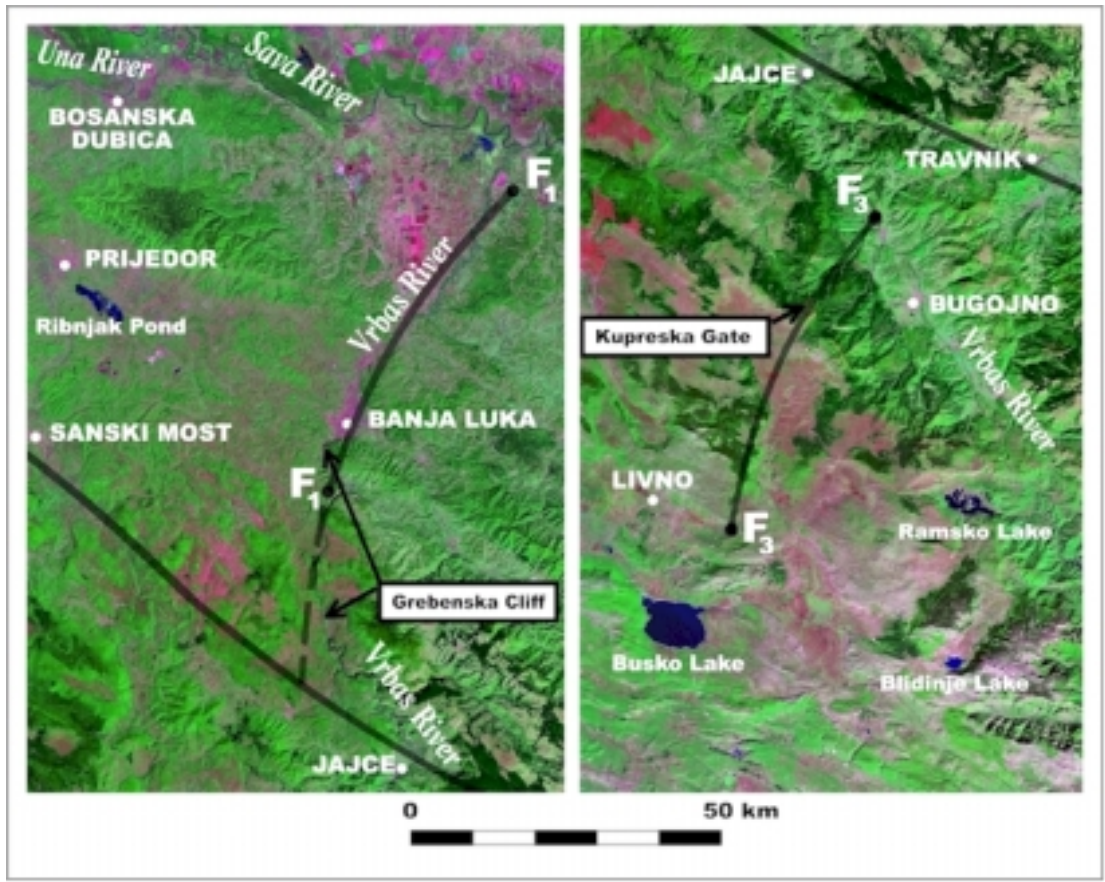

**Fig. 3** Traces of the Gradiška Fault ($F_1F_1$) – left panel, and Livno Fault ($F_3F_3$) – right panel, as parts of the region's largest, Sarajevo Fault $F_0F_0$. LANDSAT 2000 overlay.

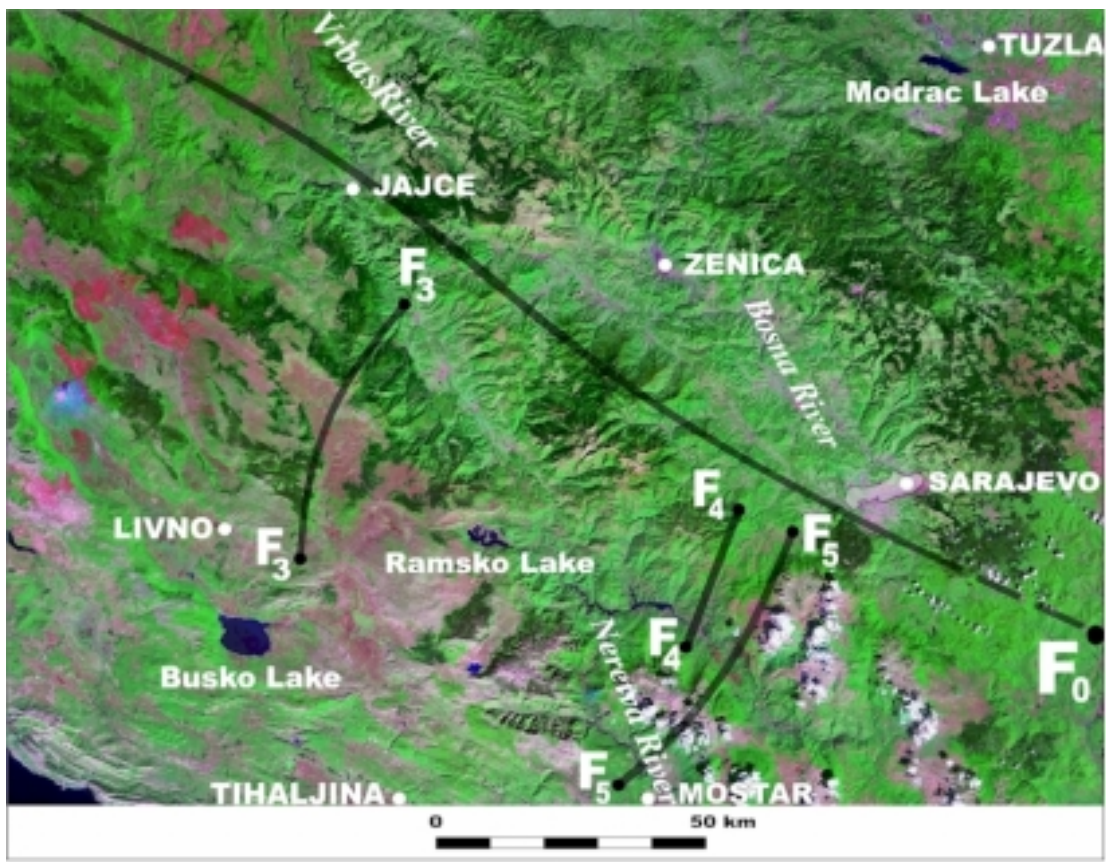

**Fig. 4** Traces of the Livno Fault ($F_3F_3$) and Mostar Fault ($F_5F_5$) as parts of the region's largest, Sarajevo Fault $F_0F_0$. LANDSAT 2000 overlay.

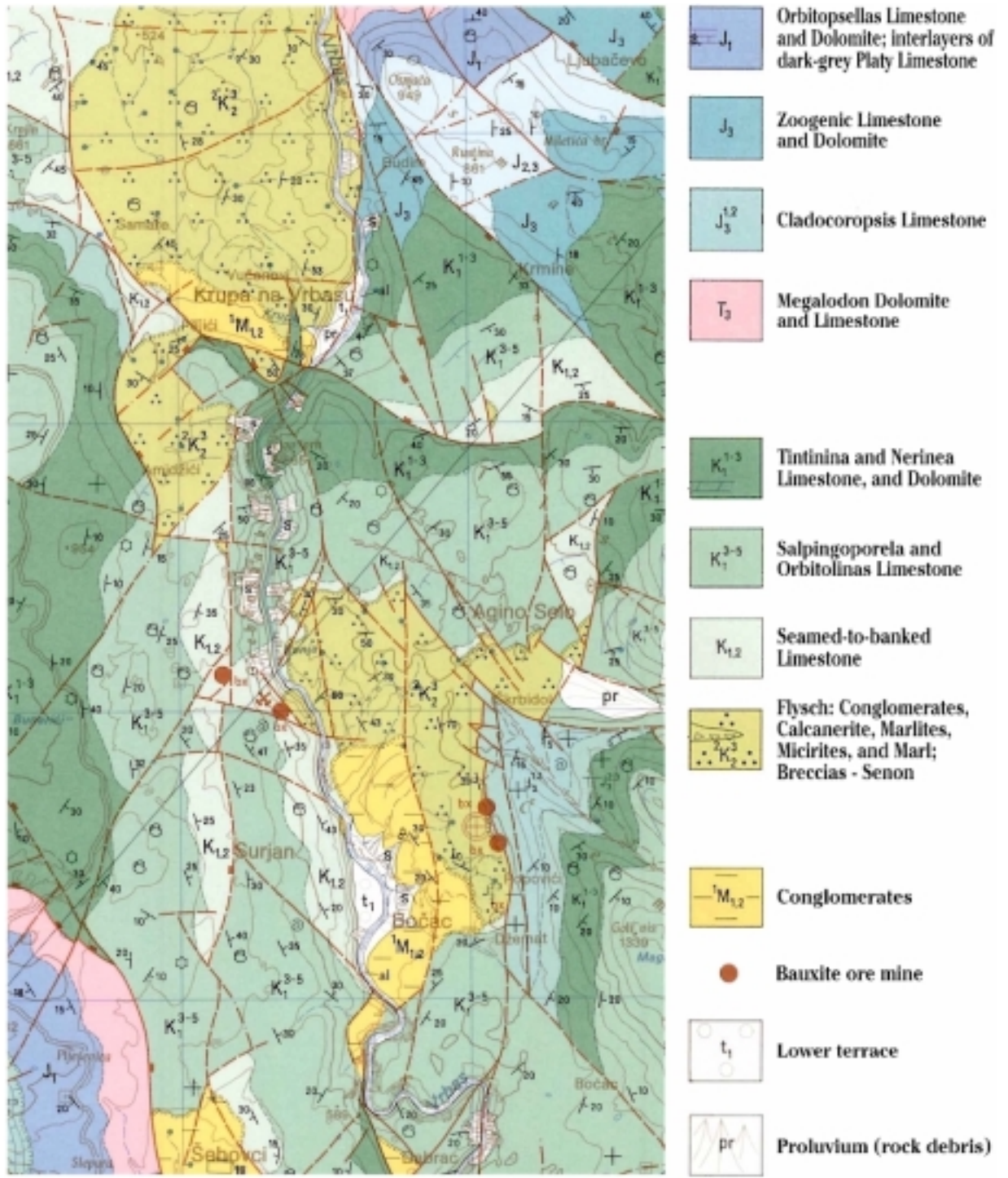

**Fig. 5**   Geological map of Grebenska Cliff (Marinković and Ahac, 1975).